\documentclass[twocolumn]{aastex63}

\newcommand{\ms}{m~s$^{-1}$}
\newcommand{\PSUAA}{Department of Astronomy \& Astrophysics, 525 Davey Laboratory, Pennsylvania State University, University Park, PA, 16802, USA}
\newcommand{\PSUCEHW}{Center for Exoplanets and Habitable Worlds, 525 Davey Laboratory, Pennsylvania State University, University Park, PA, 16802, USA}
\newcommand{\UA}{Steward Observatory, The University of Arizona, 933 N.\ Cherry Ave, Tucson, AZ 85721, USA}
\newcommand{\Penn}{Department of Physics and Astronomy, University of Pennsylvania, 209 S 33rd St, Philadelphia, PA 19104, USA}

\newcommand{\GoddardESAL}{Exoplanets and Stellar Astrophysics Laboratory, NASA Goddard Space Flight Center, Greenbelt, MD 20771, USA}

\newcommand{\UW}{Department of Astronomy, University of Wisconsin, Madison, WI 53706, USA}
\newcommand{\Macquarie}{Department of Physics and Astronomy, Macquarie University, Balaclava Road, North Ryde, NSW 2109, Australia }

\newcommand{\JPL}{Jet Propulsion Laboratory, California Institute of Technology, 4800 Oak Grove Drive, Pasadena, California 91109}
\newcommand{\MIT}{Kavli Institute for Astrophysics and Space Research, Massachusetts Institute of Technology, Cambridge, MA, USA}
\newcommand{\UCI}{Department of Physics \& Astronomy, The University of California, Irvine, Irvine, CA 92697, USA}
\newcommand{\Carleton}{Carleton College, One North College St., Northfield, MN 55057, USA}

\newcommand{\NESSF}{NASA Earth and Space Science Fellow}
\newcommand{\Sagan}{NASA Sagan Fellow}
\newcommand{\NOAO}{NSF's National Optical-Infrared Astronomy Research Laboratory, Tucson, AZ 85726, USA}

\usepackage{soul}

\accepted{February 18, 2020}
\submitjournal{AJ}

\shorttitle{Solar Contamination}
\shortauthors{Roy et al.}

\graphicspath{{./}{figures/}}

\begin{document}

\title{Solar Contamination in Extreme Precision Radial Velocity Measurements: \\Deleterious Effects and Prospects for Mitigation}
\correspondingauthor{Arpita Roy}
\email{aroy@caltech.edu}

\author[0000-0001-8127-5775]{Arpita Roy}
\affiliation{Department of Astronomy, California Institute of Technology, Pasadena, CA 91125, USA}
\affil{Robert A. Millikan Postdoctoral Fellow}
\affil{\PSUAA}
\affil{\PSUCEHW}

\author[0000-0003-1312-9391]{Sam Halverson}
\affil{\JPL}
\affil{\MIT}
\affil{\Sagan}

\author[0000-0001-9596-7983]{Suvrath Mahadevan}
\affil{\PSUAA}
\affil{\PSUCEHW}

\author[0000-0001-7409-5688]{Gudmundur Stefansson}
\affil{\PSUAA}
\affil{\PSUCEHW}
\affil{\NESSF}

\author[0000-0002-0048-2586]{Andrew Monson}
\affil{\PSUAA}
\affil{\PSUCEHW}

\author[0000-0002-9632-9382]{Sarah E.\ Logsdon}
\affil{\NOAO}

\author[0000-0003-4384-7220]{Chad F.\ Bender}
\affil{\UA}

\author[0000-0002-6096-1749]{Cullen H.\ Blake}
\affil{\Penn}

\author{Eli Golub}
\affil{\NOAO}

\author{Arvind Gupta}
\affil{\PSUAA}
\affil{\PSUCEHW}

\author{Kurt P.\ Jaehnig}
\affil{\UW}

\author[0000-0001-8401-4300]{Shubham Kanodia}
\affil{\PSUAA}
\affil{\PSUCEHW}

\author[0000-0001-6909-3856]{Kyle Kaplan}
\affil{\UA}

\author[0000-0003-0241-8956]{Michael W.\ McElwain}
\affil{\GoddardESAL} 

\author[0000-0001-8720-5612]{Joe P.\ Ninan}
\affil{\PSUAA}
\affil{\PSUCEHW}

\author{Jayadev Rajagopal}
\affil{\NOAO}

\author[0000-0003-0149-9678]{Paul Robertson}
\affil{\UCI}

\author[0000-0002-4046-987X]{Christian Schwab}
\affil{\Macquarie}

\author[0000-0002-4788-8858]{Ryan C. Terrien}
\affil{\Carleton}

\author[0000-0002-6937-9034]{Sharon Xuesong Wang}
\affiliation{The Observatories of the Carnegie Institution of Washington, 813 Santa Barbara Street, Pasadena, CA 91101, USA}

\author{Marsha J.\ Wolf}
\affil{\UW}

\author[0000-0001-6160-5888]{Jason T.\ Wright}
\affil{\PSUAA}
\affil{\PSUCEHW}

\begin{abstract}

Solar contamination, due to moonlight and atmospheric scattering of sunlight, can cause systematic errors in stellar radial velocity (RV) measurements that significantly detract from the $\sim$10~c{\ms} sensitivity required for the detection and characterization of terrestrial exoplanets in or near Habitable Zones of Sun-like stars. The addition of low-level spectral contamination at variable effective velocity offsets introduces systematic noise when measuring velocities using classical mask-based or template-based cross-correlation techniques. Here we present simulations estimating the range of RV measurement error induced by uncorrected scattered sunlight contamination. We explore potential correction techniques, using both simultaneous spectrometer sky fibers and broadband imaging via coherent fiber imaging bundles, that could reliably reduce this source of error to below the photon-noise limit of typical stellar observations. We discuss the limitations of these simulations, the underlying assumptions, and mitigation mechanisms. We also present and discuss the components designed and built into the NEID precision RV instrument for the WIYN 3.5m telescope, to serve as an ongoing resource for the community to explore and evaluate correction techniques. We emphasize that while ``bright time'' has been traditionally adequate for RV science, the goal of 10~c\ms\ precision, on the most interesting exoplanetary systems may necessitate access to darker skies for these next-generation instruments. \end{abstract}

\keywords{Radial velocity (1332), Exoplanet astronomy (486), High resolution spectroscopy (2096), Sky brightness (1462), Astronomy data analysis (1858)}

\section{Introduction} \label{sec:intro}

The field of radial velocity (RV) exoplanet detection is engaged in an ongoing battle for ever higher precision measurements, motivated by the urge to detect Earth-mass planets in the Habitable Zone \citep{Kopparapu:2013}. Such terrestrial mass planets orbiting Sun-like stars induce Doppler reflex signals of $\sim$10 cm s$^{-1}$, which nominally defines the precision goal for many next generation instruments and surveys. This realm of ``extreme" precision spectroscopy requires substantial advances in both instrument design and analysis techniques \citep{Fischer:2016}, since previously negligible sources of error become significant at these levels of scrutiny. A  number of heretofore uncharacterized phenomena that affect spectral line profiles must now be extensively studied and mitigated. Several next-generation extreme precision spectrographs \citep[see][for a complete list]{Wright:2017}, including the GMT Consortium Large Earth Finder \citep[G-CLEF;][]{Podgorski:2014}, NEID \citep{Halverson:2016}, and the Keck Planet Finder \citep[KPF;][]{Gibson:2018} are thus being built with a bottom-up systems engineering approach based on comprehensive RV error budgets, which can be valuable tools to predict instrument performance. 

\begin{figure*}[ht]
    \begin{center}
        \includegraphics[width=0.7\textwidth]{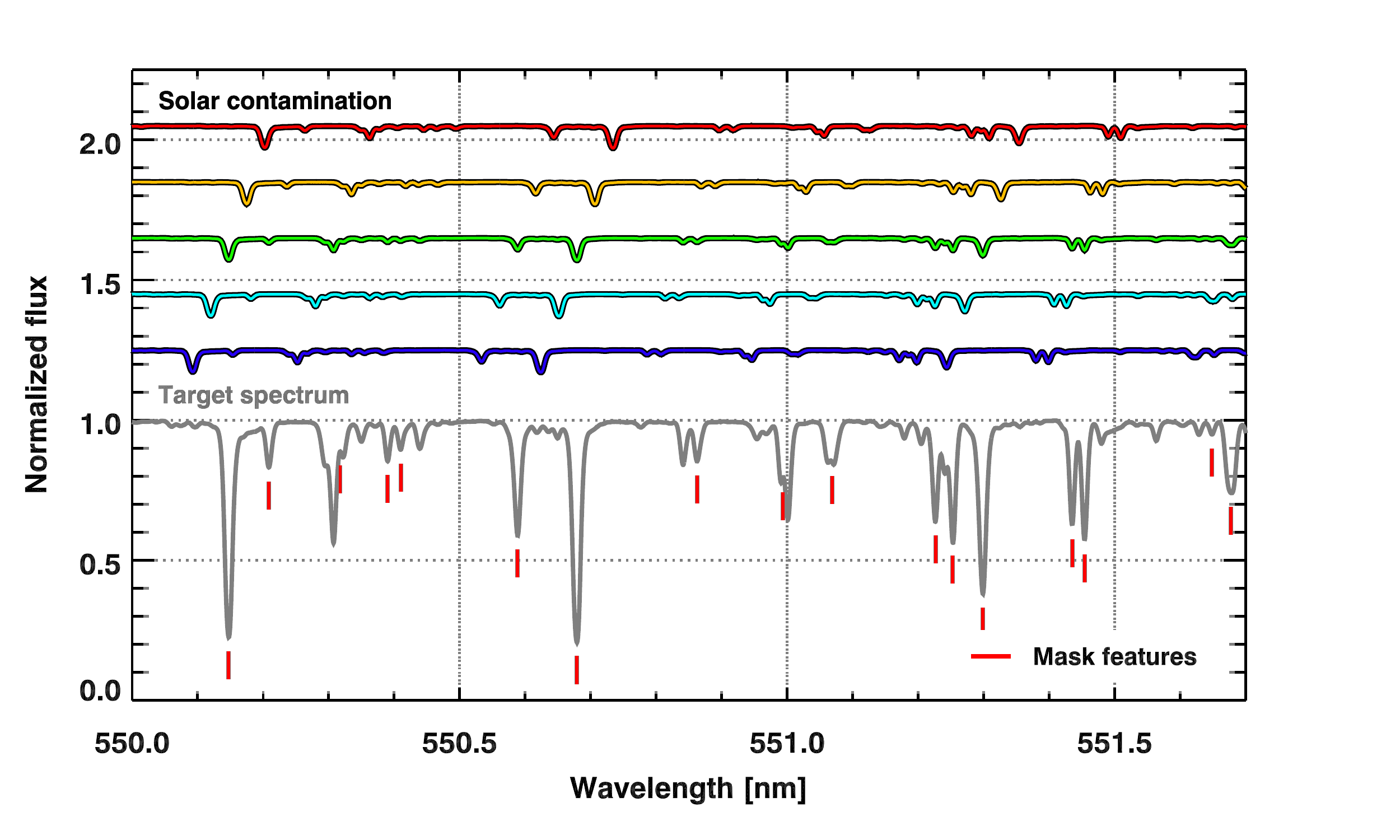}
        \includegraphics[width=0.7\textwidth]{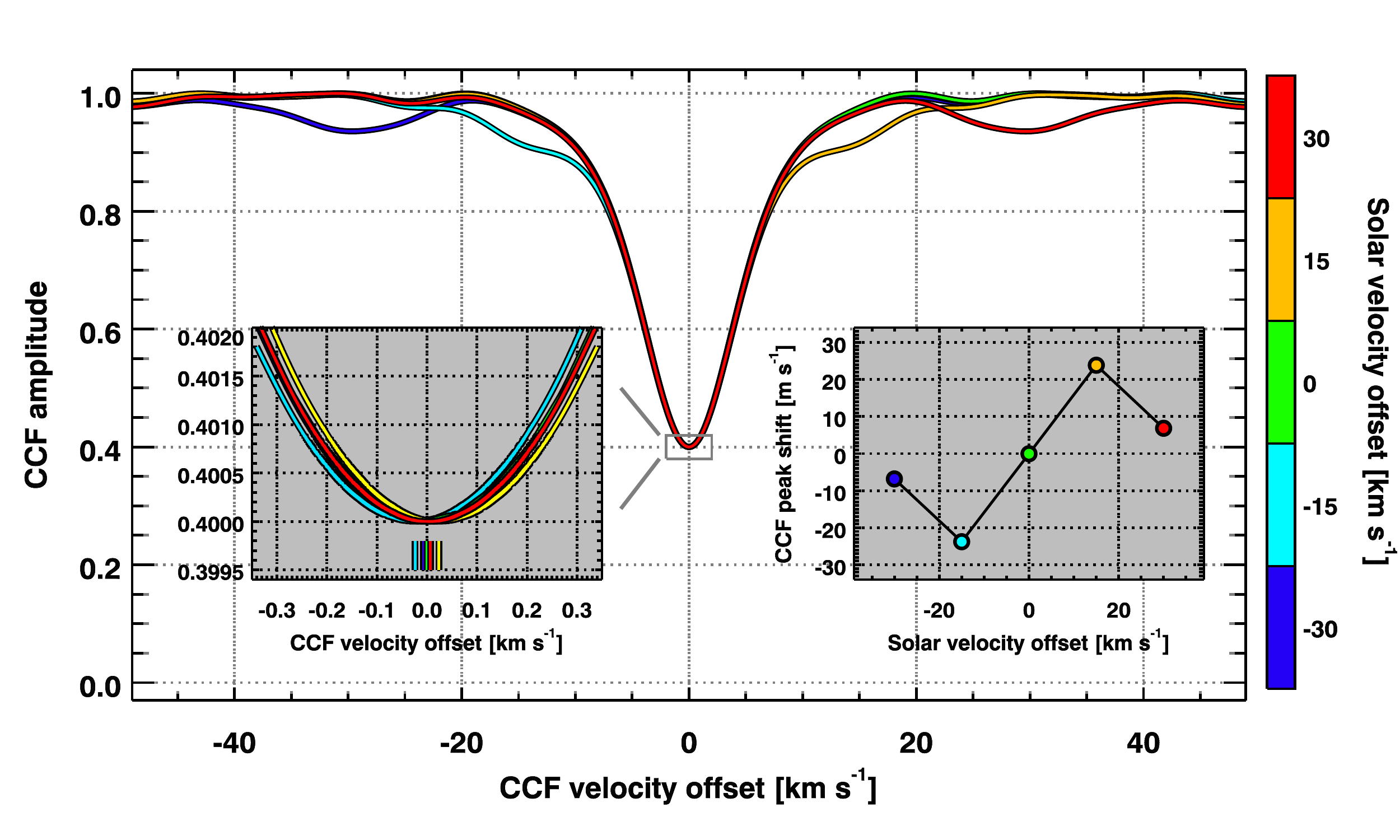}
        \caption{Toy model for illustration of `peak-pulling' of the measured cross-correlation function (CCF) due to scattered solar contamination. \textit{Top:} A relatively low amplitude `sky' solar spectrum (colored curves) is added to a synthetic G2 dwarf `target' spectrum (gray) at different absolute velocity offsets. In this illustration, we exaggerate `sky' to be 10\% of the target brightness to highlight the impact. Spectral lines used to construct the aggregate CCF are highlighted in red. \textit{Bottom:} The effective velocity offset of the combined spectrum relative to the target spectrum is calculated by fitting a Gaussian to the peak of the CCF. As the cross-correlation mask encounters solar features, spurious peaks are introduced into the CCF that can lead to significant velocity offsets if left uncorrected (see inset figures).}
        \label{fig:ccf_ex}
    \end{center}
\end{figure*}

\begin{figure*}[ht]
    \begin{center}
        \includegraphics[width=1\textwidth]{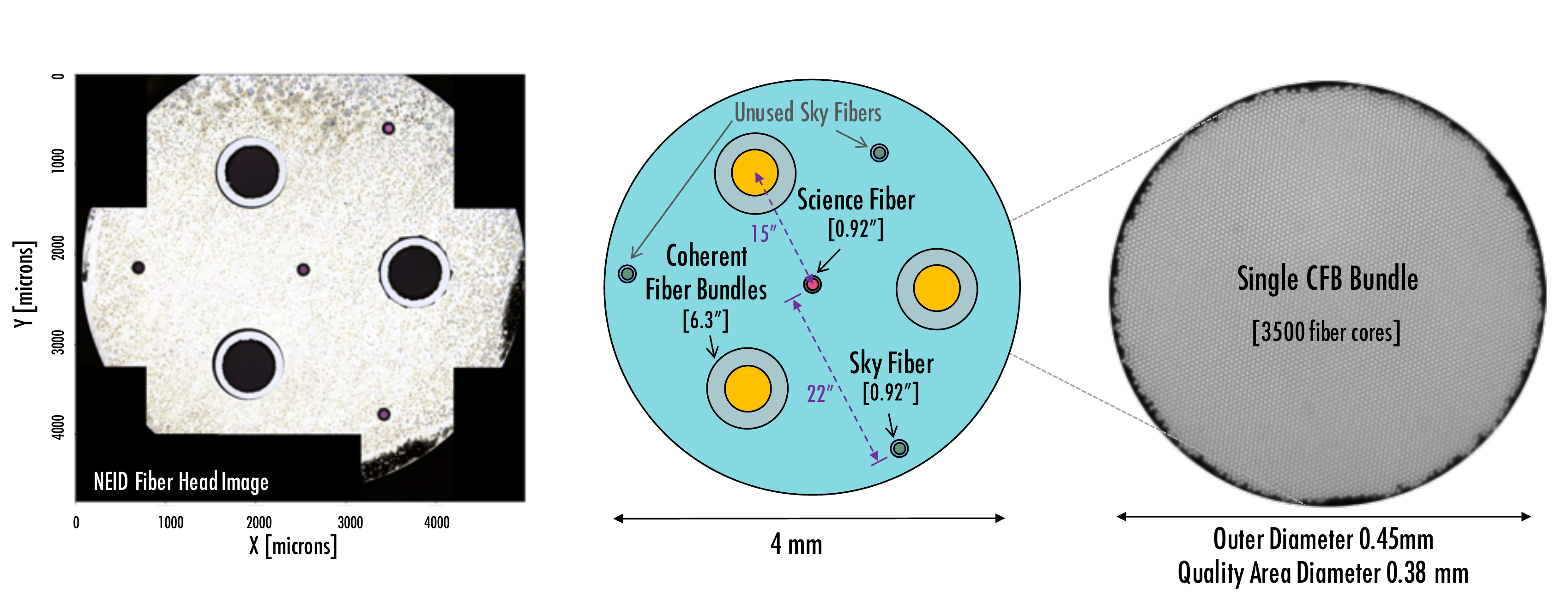}
        \caption{NEID high resolution (HR) mode port adapter fiber head that collects light from the telescope for transportation to the spectrometer. \textit{Left:} Image of the as-built fiber head being installed with NEID, taken through a Zygo Optical Profilometer which uses white light interferometry to obtain a high precision image. \textit{Middle:} Diagram showing the relative placement of the coherent fiber bundles (CFBs) around the science fiber, as well as the position of the sky fiber (unused sky fibers provide redundancy). \textit{Right:} Each CFB is composed of several thousand fiber cores; 3,500 for NEID specifically. Magnified CFB image from the manufacturer (SCHOTT).}
        \label{fig:fiber_head}
    \end{center}
\end{figure*}

\begin{figure*}[ht]
    \begin{center}
        \includegraphics[width=0.8\textwidth]{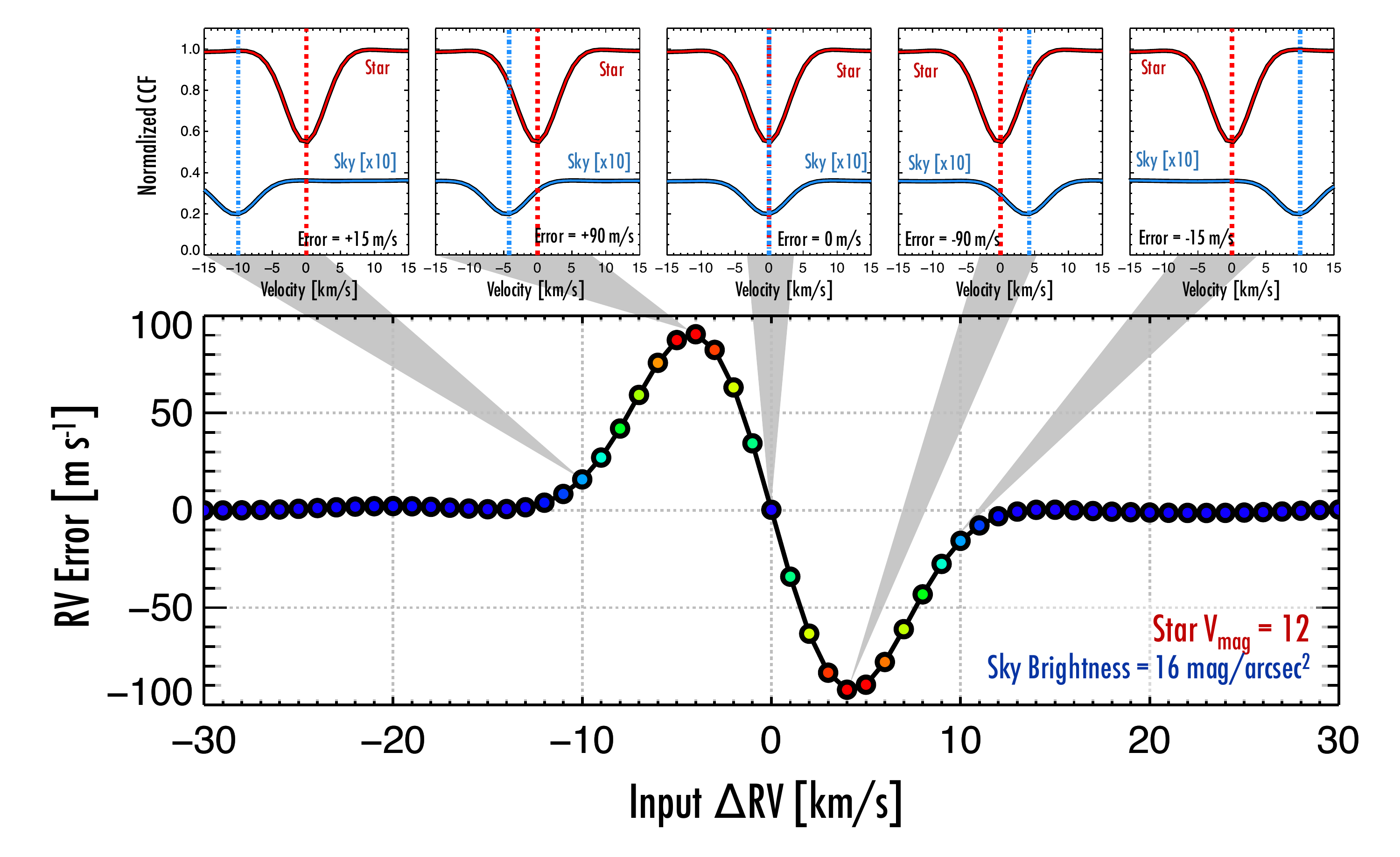}
        \caption{Radial velocity error caused by solar contamination, as a function of velocity separation, for a given stellar V$_{\rm mag}$ and sky brightness. \textit{Top panels:} The contaminant sky spectrum is shifted across the full range of possible barycentric velocities ($\pm 30$ k\ms) against the target star spectrum in its rest frame. Here we show the resulting CCFs for star and sky; the sky CCF is multiplied by 10 for visibility. \textit{Bottom:} The RV error is low when the star is well separated from the sky spectrum in velocity space. RV error is highest when the separation between star and sky is $\sim$4~k\ms for NEID, as the contaminant peak passes through the wings of the primary peak. If the star and sky spectrum are perfectly aligned, there is essentially no RV impact (although note that any scrutiny of line shape metrics might be affected).}
        \label{fig:rvrange}
    \end{center}
\end{figure*}

Spectral contamination remains an area of grave concern in formulating RV error estimates since it significantly detracts from our ability to measure intrinsic stellar line profiles. \citep{Pepe:2008}. The worst case of contamination occurs when the superimposed light contains spectral features, which can both pollute individual line shapes and induce a spurious velocity signal across the ensemble of lines. This is exacerbated by the fact that relative shifts between contaminant and source spectra are often introduced by the motion of the star and Earth-based observatory over time. In the case of background sky brightness interfering with RV measurements, there are no hardware solutions to entirely eliminate this effect for seeing-limited instruments. In fact, the magnitude of the contaminant must often be diagnosed from the same observations that need to be corrected for the effect. Mitigating this by trying to observe only in dark time is possible but impractical, given competition with other sub-fields like extragalactic and cosmological studies.

Blended light from companions or inconvenient background objects are known to be capable of producing deceptive false positives, deserving careful treatment \citep[e.g.][]{Wright:2013}. Unlike the incidental nature of these contaminants, however, we focus here on scattered or reflected sunlight, which is essentially always present in observed spectra from ground-based instruments at some level. The magnitude of this contamination depends on several factors including sky brightness, target-moon separation, lunar phase, ecliptic latitude, zenith angle, and phase of the solar cycle \citep{Krisciunas:1997}. Thus far, this error source has largely been mitigated by avoiding (i) twilight observations, (ii) observations during full moon or when the target is close to the moon, and (iii) observations during cloudy skies or cirrus \citep{Pepe:2008, Seager:2010a}. Ensuring that the sky background is faint ($>$7-10 magnitudes fainter as a rule of thumb) has been largely adequate for current instruments operating in the $\ge$1~m~s$^{-1}$ precision regime \citep{Pepe:2008}.

The feasibility of performing solar contamination correction is yet to be explored for Doppler measurements at the 10~c{\ms} level. The presence of scattered sunlight in observed spectra can be a two-fold source of error in weighted mask-based \citep{Pepe:2002} or high signal-to-noise ratio (SNR) template-based cross correlation for RV measurement, affecting both `peak pulling' and the introduction of more complex and time variable structure in the cross-correlation function (CCF) or $\chi^2$ space (Figure \ref{fig:ccf_ex}). 

Here we carefully consider the impact of solar light contamination on RV measurements and explore the mitigation possible with a simultaneous sky fiber. Several upcoming instruments, e.g., NEID \citep{Schwab:2016} and the Keck Planet Finder \citep{Gibson:2018}, include a dedicated sky fiber for the correction of telluric absorption and emission lines, and the removal of scattered sunlight. However, the sky fiber is highly spectrally dispersed in these instruments, making faint levels of solar contamination difficult to measure even when there is a consequential impact on RV precision.

For cases of very faint sky with low signal-to-noise ratio (SNR) in the sky fiber, we examine the viability of substituting the sky fiber with broadband sky images from coherent fiber bundles (CFBs). CFBs are included in the NEID fiber head primarily for target acquisition purposes, but present an interesting alternative path for sky brightness measurements. We attempt to keep our simulation fairly general across seeing-limited next-generation spectrographs. Where specification is necessary, we base our parameters on the NEID spectrograph, currently being commissioned on the 3.5m WIYN telescope at Kitt Peak. As a result, these calculations are readily adaptable for other instruments with adjustments for telescope aperture, fiber size and shape, instrument bandpass and resolution, and expected sky conditions.

We examine the impact of sky brightness on RV measurements in \S\ref{sec:impact}, including descriptions of our simulations, measurement technique, and the magnitude of the uncorrected error. We present three correction strategies in \S\ref{sec:correction}, and detail their inclusion in the NEID design in \S\ref{sec:neidcfb}. We discuss our results and some nuanced instrument-specific effects in \S\ref{sec:discussion}, and conclude in \S\ref{sec:conclusion}.

\section{Impact of Sky Brightness on RV Measurements}
\label{sec:impact} 

Here we set out to predict the deleterious effects of solar contamination on RV precision and prepare our mitigation techniques that leverage the sky fiber, in close anticipation of detailed tests on NEID. For this exercise, it is important to create realistic spectra that accurately model the absolute flux and noise inherent to different levels of sky brightness. Instruments like NEID use a dedicated sky fiber to directly sample the solar contamination spectrum at a resolution identical to the target spectrum. Since the science and sky fibers are the same size, and at relatively small separations (22\arcsec on-sky in the case of NEID, Figure \ref{fig:fiber_head}), they should contain very similar amounts of sky flux. Both fibers should also have comparable throughput, or be flux calibrated to high degrees of confidence. This similarity is quantified by measurement on NEID, but must be verified for individual instruments.

\subsection{Simulating Science and Sky Fiber Spectra}
We simulate both `sky' and `star' spectra using a synthetic high resolution solar spectrum (R$\sim$500,000; T$_{\mathrm{eff}}$=5800K; log(g)=4.5; [Fe/H]=0.0) computed with the PHOENIX stellar atmosphere models \citep{Husser:2013}. We deliberately opt to use the same synthetic spectrum for both sources (i.e., the case when you are observing a Sun-like star) since Doppler errors are maximized when the contaminant is similar to the observed target, thus rendering a worst case scenario. To ensure accurate photon counts corresponding to different stellar magnitude levels, we scale the model spectrum using the absolute flux distribution of solar analogs based on spectrophotometry from the Hubble Space Telescope Faint Object Spectrograph \citep{Colina:1997}. The spectrum is then attenuated for atmospheric extinction at Kitt Peak, and converted to flux incident on the 3.5m WIYN telescope aperture. 

Wavelength dependent seeing losses through an octagonal fiber are calculated using a median WIYN PSF with pODI (the precursor to ODI, the One Degree Imager at WIYN) in the r-band provided by NOAO, ahead of extensive characterization with NEID. This PSF is highly non-Gaussian and shows an extended aureole, as expected from Kolmogorov theory, and we model it with a 3-Moffat function as recommended by \cite{Racine:1996}. To this we apply the complete NEID spectrometer throughput model, which accounts for all subsequent losses in the system.  

These steps are repeated for the `sky' spectrum, after converting a desired sky brightness (in mag/arcsec$^2$) to a V-band magnitude for the 0.92'' NEID fibers for the high resolution mode; keeping in mind that the sky fiber does not suffer seeing losses due to the uniform illumination on the fiber face. While the sky fiber is immune to atmospheric dispersion effects, note that it is important to have considered performance requirements on the atmospheric dispersion corrector (ADC) to trace photon incidence rates through the science fiber, and ensure reliable chromatic count rates across the instrument bandpass \citep{Wehbe:2019, Logsdon:2018}.

The synthetic spectra are convolved with a Gaussian kernel to match the NEID instrument spectral resolution (R = 100,000), and binned to the approximate point-spread function sampling (5 pixels FWHM). The spectra are further distributed into 68 echelle orders (3800 -- 6800 \AA) based on the predicted NEID order mapping. Even though NEID extends to 9300 \AA, we limit ourselves to $\leqslant$7000 \AA~because this is the region that contains most of the Doppler information content for Sun-like stars, and to which our best cross-correlation masks are currently limited (from instruments classically calibrated with ThAr lamps). Each order is multiplied by an artificial blaze function derived from measured grating response curves of an R4 echelle. This ensures that the edges of the orders are realistically downweighted with respect to the centers, and the associated noise levels are generated correctly. To calculate the photons in each pixel of the extracted spectrum, we assume the 2-D flux profile across the fiber trace is collapsed in the cross-dispersion direction (fiber image has a width of $\sim$ 5 pixels) and scale by a standard exposure time of 20~minutes for all simulations. Poissonian photon noise and Gaussian read noise (4$\mathrm{e^-}$ per pixel based on the performance of the NEID CCD, allotting gain = 1) are added to both the target and sky spectra. The inclusion of both photon noise and read noise is critical, since at R$\sim$100,000 the photon flux rate in the sky fiber quickly becomes comparable to the read noise as sky brightness decreases. We simulate the science fiber by adding the `scattered sunlight' and the stellar spectra at realistically varying velocity separations based on the extent of barycentric motion ($\pm$30~k\ms), to mimic a range of observational epochs. The sky fiber contains only the scattered sunlight component.

\subsection{Measuring RV Errors in Simulated Spectra}
\label{subsec:measure} 

The simulated science and sky fiber spectra are passed through our RV analysis code, which has been extensively vetted at the $<$1~\ms~~level on both PARAS and HARPS data \citep{Roy:2016}. Cross-correlation functions (CCFs) are calculated using a standard stellar mask-based technique \citep{Baranne:1996, Pepe:2002, Chakraborty:2014} for each order, then summed to form an aggregate CCF. Effective velocity offsets are measured by fitting a Gaussian to the peak of the aggregate CCF. We fit $\pm$20~k\ms~around the expected peak value, similar to our implementation in the PARAS pipeline, since fitting smaller intervals around the peak increases RV error. 

Note that echelle blaze functions are not removed during processing since they are static in this idealized case, and preserve the true photon counts associated with noise generation. This also means that there is no necessity for order reweighting in this simulation. In reality, of course, changes in illumination due to clouds, seeing, telescope focus drifts, airmass, and other observing parameters, render a slightly variable blaze response and order weighting that must be either removed or corrected during spectral processing \citep[e.g][]{Anglada-Escude:2012}.

Beyond the proven performance of our code on real data from current RV instruments, here we want to be sensitive to much smaller errors corresponding to the precision of next generation spectrographs. Hence, we begin by checking our algorithmic error on ``perfect" noiseless simulations. We find that we can both introduce RV offsets and recover these injected velocities at the $<$10$^{-4}$~c\ms~~level. This ensures that we can robustly simulate and investigate contamination errors at the $\ll10$~c\ms~~level of interest, while treating algorithmic error as negligible.

\begin{figure*}[ht]
    \begin{center}
        \includegraphics[width=1\textwidth]{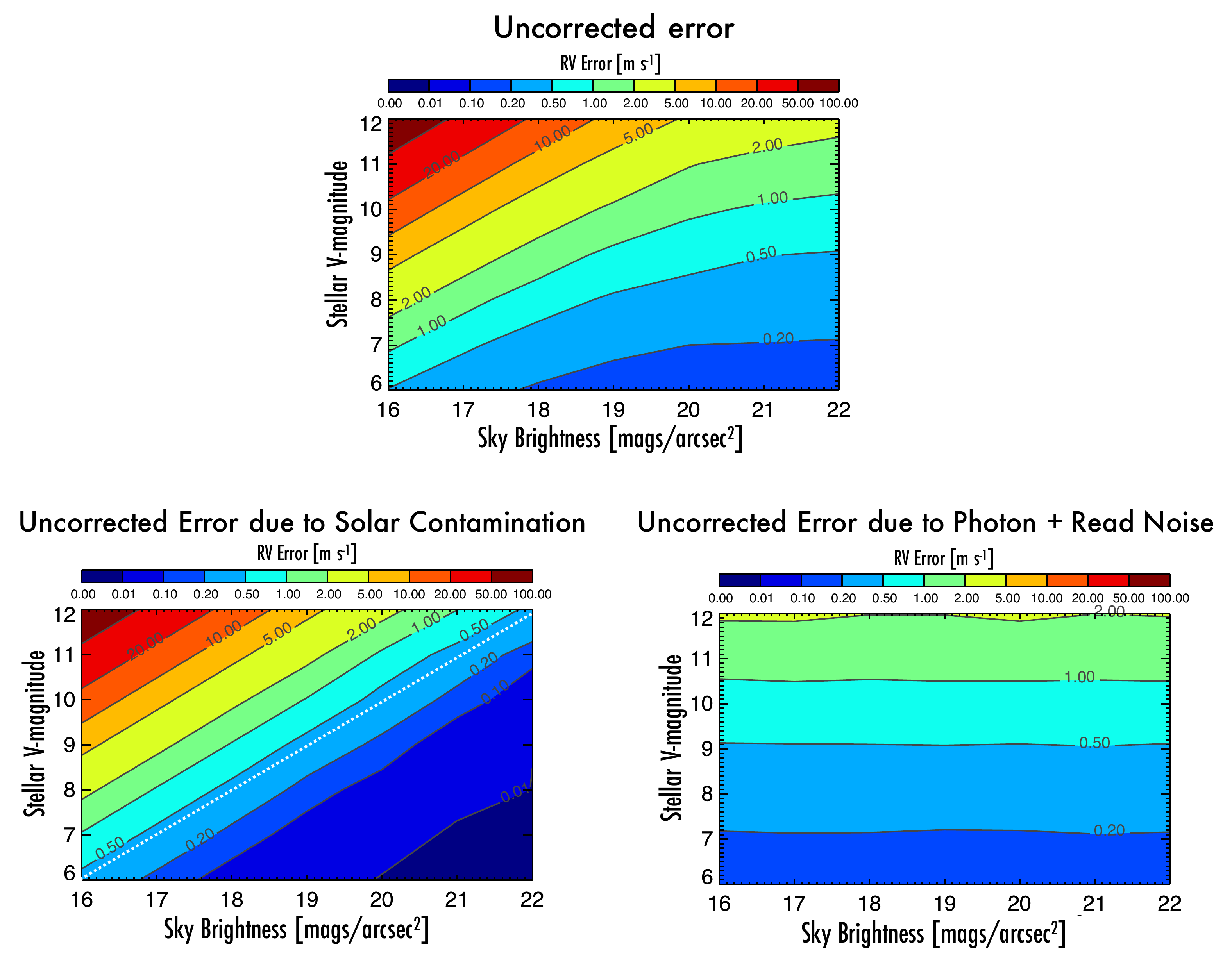}
        \caption{Uncorrected radial velocity errors introduced by the presence of contaminating moonlight or scattered sunlight, for a range of stellar V-band magnitudes and sky brightness levels. Note that these are worst case errors, calculated at the velocity separation between the star and the sky that causes maximum RV displacement. \textit{Top:} Simulations with realistic noise --- including photon noise, read noise, and solar contamination, showing the full measured RV error. The full uncorrected error is separated into two categories in the bottom panels. \textit{Bottom Left:} Isolating the error due to solar contamination. The dashed white line illustrates the error incurred following the traditional practice of observing when sky is $\sim$10 magnitudes fainter than the target. \textit{Bottom Right:} Contribution of all other error sources, illustrating the achievable noise limit for each stellar brightness.}
        \label{fig:uncorrected}
    \end{center}
\end{figure*}

\subsection{Uncorrected Solar Contamination Errors}
\label{subsec:uncorrected} 

Our set of simulated scenarios encompass stars with V$_{\rm {mag}}$= 6 - 12, and sky brightnesses from 16~--~22~mag/arcsec$^2$. This is based on typical targets for extreme precision RV measurements on NEID, and the range of conditions at Kitt Peak National Observatory (KPNO). KPNO dark time sky brightness is V$\sim$21.45~mags/arcsec$^2$ at zenith, but increases rapidly to V$\sim$20.9~mag/arcsec$^2$ (airmass=2) in the direction of Tucson \citep{Massey:2000, Neugent:2010}. During ``grey time'' at the WIYN site ($>$3 days from full moon), additional moon contamination can add up to 1.8~mag/arcsec$^2$ in the V-band \citep{a1987}. We simulate this full range of overall sky brightness, even though some of it is certainly from airglow and city lights, which interacts with the target spectrum differently than the scattered sunlight spectrum. 

The contaminating spectra are shifted in velocity relative to the target spectrum to span the full range of potential velocity offsets due to the barycentric motion of the Earth, approximately $\pm$30~k\ms. As shown in  Figure \ref{fig:rvrange}, the contaminant affects maximum peak pulling (and correspondingly RV error), when it begins to blend into the wings of the primary CCF peak. The separation that causes the highest RV error is set principally by the resolution of the instrument, and to a lesser degree by choices like mask width and the extent of the peak being fitted during RV measurement. For NEID this is at approximately $\pm$4~k\ms, close to the full-width at half maximum of the CCF. However, this is not the only velocity separation of concern, since there are also smaller ancillary peaks from the sky spectrum interacting with the target spectrum that cause errors at the few percent level of the worst case (e.g., at $\pm$20~k{\ms} in Figure \ref{fig:rvrange}), implying that contamination correction cannot be easily dismissed at any velocity separation. Since this is a decidedly non-Gaussian error, and for the sake of estimating the worst case impact, here we focus on the maximum absolute RV error for each combination of star and sky brightness (i.e., the maximum values in Figure \ref{fig:rvrange}, bottom panel). We generate 1000 realizations of spectra with realistic noise properties for each scenario.

If left uncorrected, solar contamination can cause substantial RV variations that overwhelm other aspects of the error budget. Figure \ref{fig:uncorrected} shows the maximum RV error caused by a range of sky brightnesses for typical NEID targets. The top panel shows the full uncorrected error (mean + 1-$\sigma$ uncertainty over 1000 realizations), which is the realistic limit imposed by the combined effect of solar contamination, photon noise, read noise, and the mask cross-correlation technique. We separate these errors into two categories in the bottom panel.

In the bottom right panel we isolate the error caused by solar contamination alone, which contributes the bulk of the mean error in the top panel. In the bottom left panel we show the achievable noise floor for each stellar magnitude as derived from 1000 injection recovery runs on uncontaminated stellar spectra. This encapsulates all other error sources in the simulation. It is dominated by photon and read noise, but algorithmic errors are also naturally included during RV measurement.

The traditional practice of observing when sky is $\sim$10 magnitudes fainter than the target adds contamination errors at the 0.5~\ms~~level, as illustrated by the white dashed diagonal across the bottom left plot in Figure \ref{fig:uncorrected}. This is consistent with overall operations at the 1~\ms~~level of precision. However, it is inadequate for instruments hoping to detect terrestrial-mass planets with actual RV semi-amplitudes of 10~c\ms, since solar contamination has the potential to overwhelm that limit across most of our parameter space. Without correction, a magnitude differential of $>$12 seems necessary to limit the isolated error from solar contamination to $<$10~c\ms. This becomes problematic when we consider that bright time is typically allotted for exoplanet radial velocity science. Achieving sensitivity at the 10~c\ms\ level on sky poses stringent additional requirements on observing conditions that must be be factored in during time allotment on these next-generation instruments (\S\ref{sec:conclusion}).

\begin{figure*}[t]
    \begin{center}
        \includegraphics[width=0.7\textwidth]{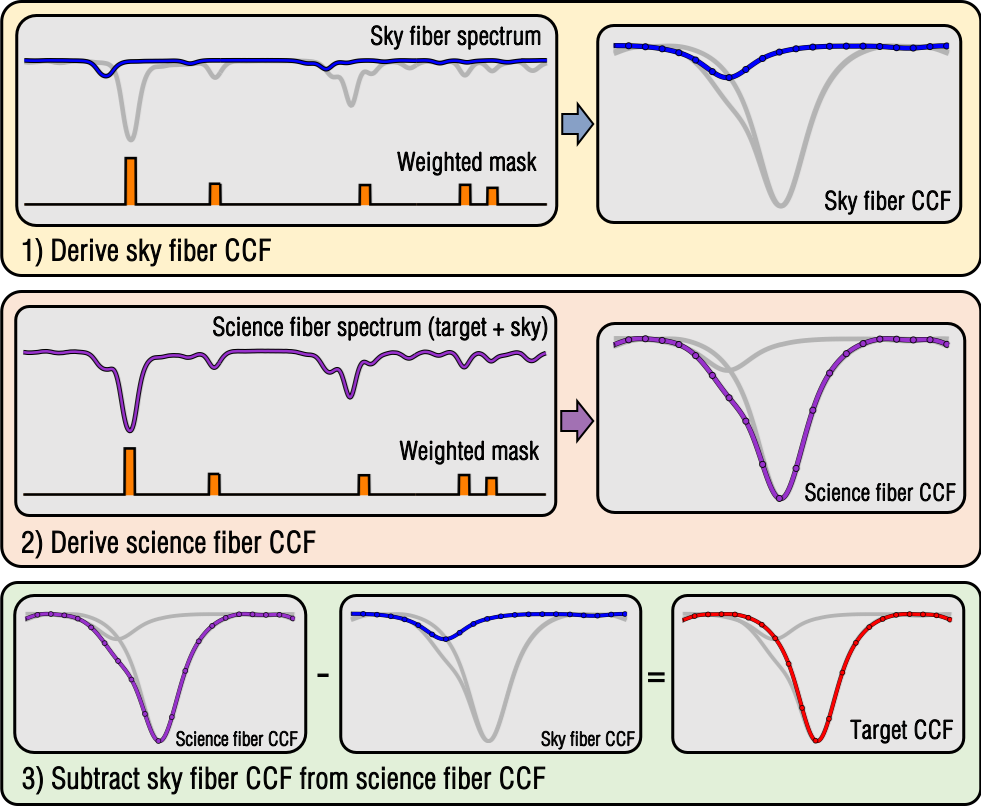}
        \caption{Overview of the direct sky fiber cross-correlation CCF subtraction technique, as described in \S\ref{subsec:direct}. The CCF of the sky fiber is computed by cross-correlation with a numerical mask (top) and subtracted from the CCF of the science fiber spectrum (middle). The final target RV is derived from the resultant corrected target CCF (bottom). This technique has the advantage of simplicity, although does inherently rely on the sky and science fiber PSFs to be similar. Note that the spectrum and CCF continuum levels are matched for illustrative purposes only -- in reality, the sky fiber has a much lower flux level.}
        \label{fig:ccf_subtract}
    \end{center}
\end{figure*}

\begin{figure*}[t]
    \begin{center}
        \includegraphics[width=1\textwidth, trim = {0 1cm 0 1.5cm}]{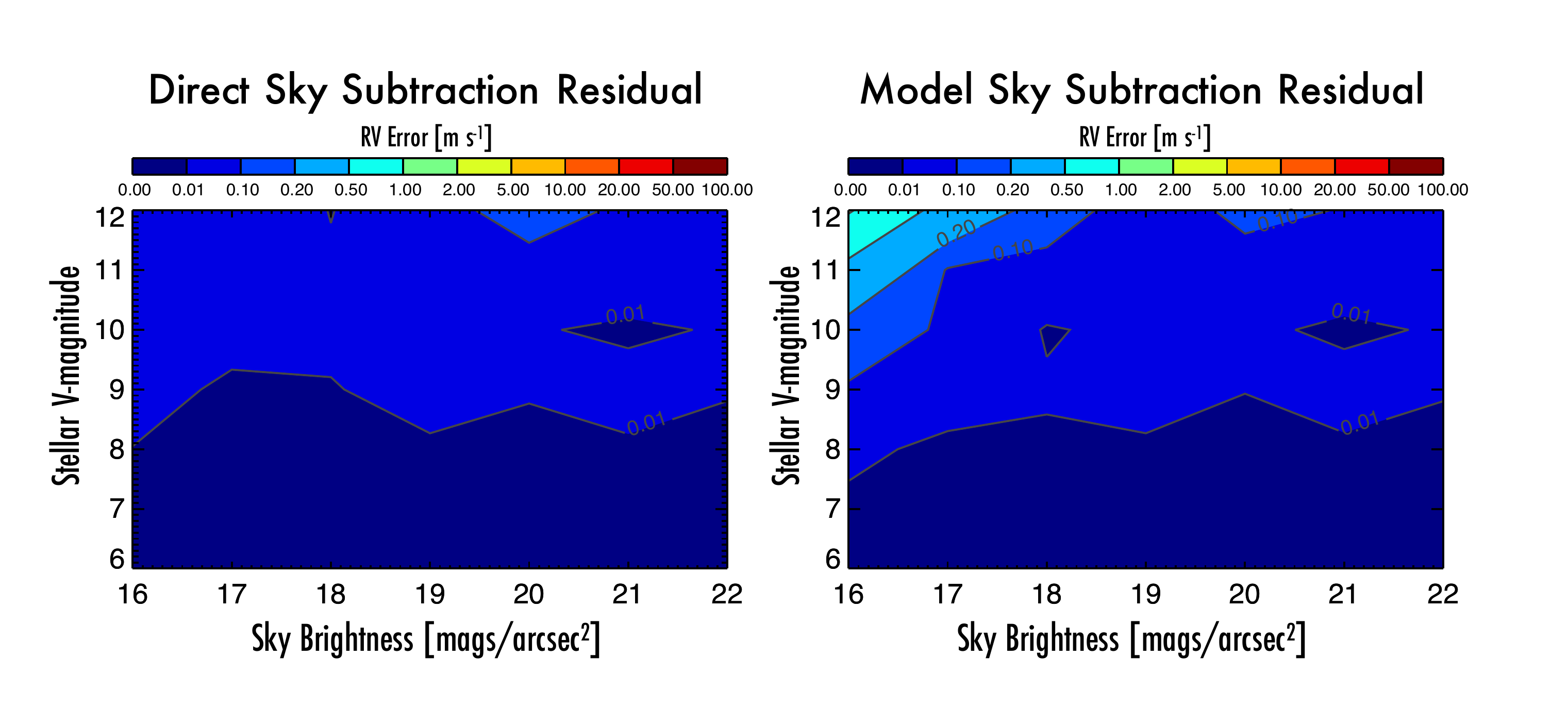}
        \caption{Residual radial velocity errors from solar contamination after applying two different correction methods. Details of the simulations are the same as in Figure \ref{fig:uncorrected}. \textit{Left:} Subtracting the sky fiber CCF directly from the science fiber CCF, for a range of stellar V-band magnitudes and sky brightness levels. \textit{Right:} Subtracting a model sky spectrum (constructed based on the flux in the simulated sky fiber) from the science fiber spectrum. These results should be compared to the uncorrected solar contamination in Figure \ref{fig:uncorrected} (bottom right), and do not include the photon or read noise contribution.}
        \label{fig:corrected}
    \end{center}
\end{figure*}

\section{Correcting for Solar Contamination}
\label{sec:correction}

Solar contamination correction will need to be an essential algorithmic part of next generation radial velocity pipelines, optimizing the use of the sky fiber and other ancillary data across the observing conditions of the telescope site. Here we explore methods to correct for this effect, the level of correction possible, and inherent limitations. 

\subsection{Direct Sky Fiber CCF Subtraction}
\label{subsec:direct}
In the field of precision RV measurements, sky correction has been previously attempted by directly subtracting the CCF of the sky fiber spectrum from the CCF of the science fiber spectrum. An overview of this technique is shown in Figure~\ref{fig:ccf_subtract}. Direct CCF subtraction is a simple but powerful technique that leverages the inclusion of a sky fiber with an identical fiber size and cross-section (and hence very similar line spread function) as the science fiber. With a real instrument, the two fibers can point at slightly different parts of the sky and have different throughputs and aberrations, necessitating a relative scaling between the CCFs before subtraction. This method has been successfully implemented for the SOPHIE spectrograph, although at precisions of 20-100~{\ms} it was mainly intended as a coarse correction of strong moonlight contamination to salvage certain observations \citep{Pollacco:2008, Barge:2008, Hebrard:2008, Bonomo:2010, Santerne:2011, Santerne:2011a}. \citet{Bonomo:2010} also note that peak pulling from moonlight contamination is immediately evident in the CCF bisectors, a method we often use to  detect spectral contamination \citep{Wright:2013} or stellar activity issues \citep{Robertson:2014} in RV measurements.

We test this correction method by subtracting the sky fiber CCF from the science fiber CCF for our simulated scenarios in Figure \ref{fig:corrected} (left). This reduces the contamination effect considerably, bringing the worst RV errors down from $\sim$100~\ms~to $\sim$10~c\ms. This residual error is due to both a shift in the mean velocity measured after CCF subtraction (residual peak pulling from sky), and an increase in the standard deviation of the measured noise distribution (due to noise in the subtracted sky fiber CCF). In fact, as sky gets darker, the sky fiber CCF becomes read noise dominated and visibly indistinguishable from noise and does not actually apply any correction. That effect, however, is obscured by the fact that contamination errors are already low when sky is dark. Since these small shifts are difficult to measure, even on simulated spectra, irregularities persist in the contours even after running 1000 realizations for each combination of star and sky. 

For the brightest NEID targets (V$_{\rm {mag}}<$8.5), this simple correction brings the errors down to $<$1~c\ms~for the full range of expected sky brightnesses at Kitt Peak. For fainter stars, this residual error can grow to $\sim$10~c\ms. Even though the residual error from solar contamination is lower than the noise floor across these observing conditions (bottom left of Figure \ref{fig:uncorrected}), it is not negligible in the context of 10~c\ms~long-term on-sky precision, and must be accounted for in a comprehensive error budget.

\begin{figure*}[t]
    \begin{center}
        \includegraphics[width=0.9\textwidth]{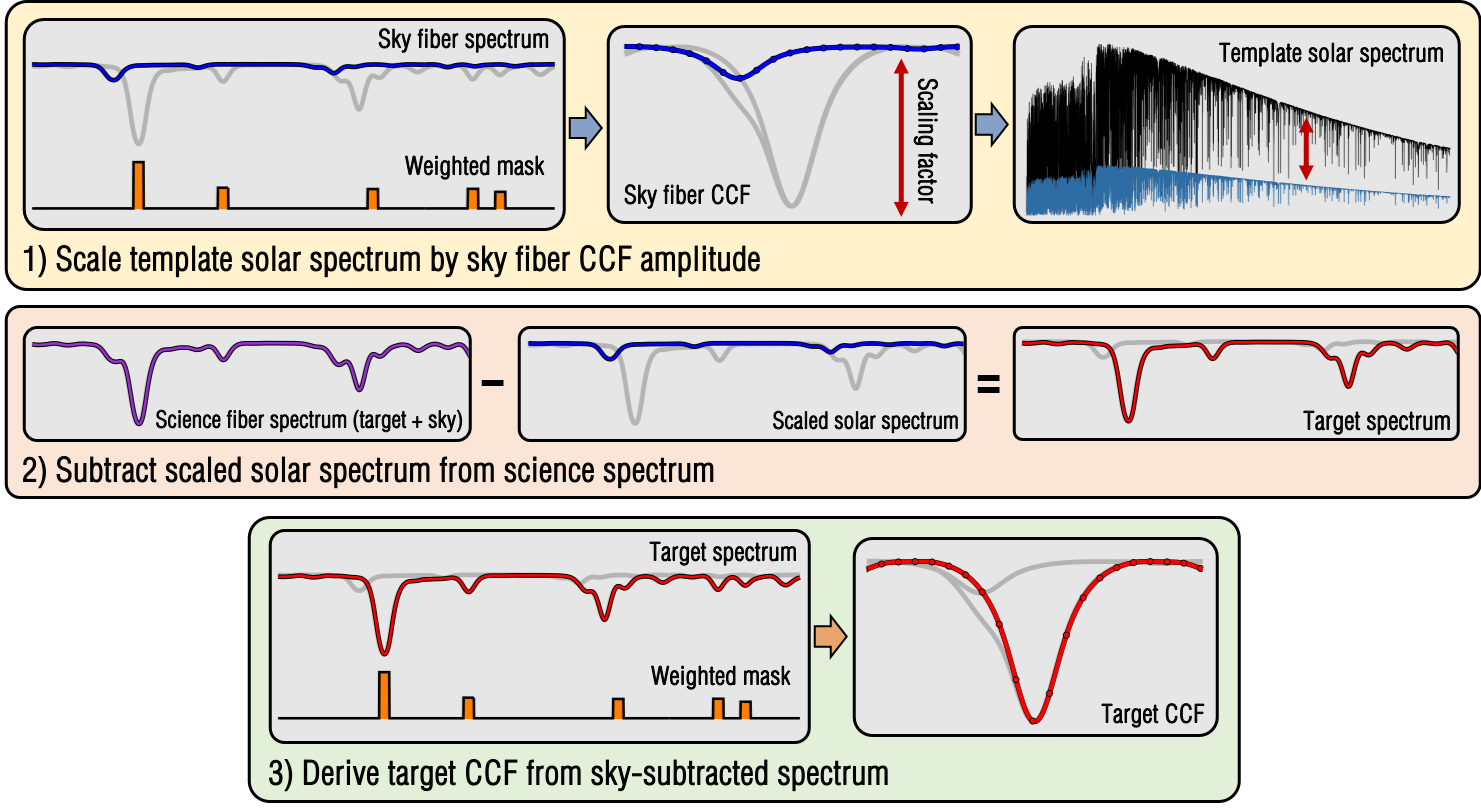}
        \caption{Overview of the model sky subtraction technique, as described in \S\ref{subsec:model}. The CCF of the sky fiber is computed using a numerical mask, and used to scale a synthetic or high SNR solar spectrum to the estimated sky brightness (top). The scaled spectrum is then subtracted from the science spectrum, leaving only the target spectrum (middle). The final target radial velocity is then computed from the target CCF (bottom). Note that the spectrum and CCF continuum levels are matched for illustrative purposes only -- in reality, the sky fiber has a much lower flux level.}
        \label{fig:ccf_scale}
    \end{center}
\end{figure*}

\subsection{Model Sky Subtraction}
\label{subsec:model}
A different method of contamination removal involves the subtraction of a noiseless (or very high SNR) model sky spectrum from the science fiber spectrum, based on the flux recorded in the sky fiber (see Figure~\ref{fig:ccf_scale}). The continuum level of the sky fiber CCF is a directly related to the absolute sky brightness:
\begin{equation}
{\rm {Sky~Brightness}} = 2.5 \times {\rm log}\bigg{(}\frac{\rm {constant}}{\rm {CCF~continuum}}\bigg{)}
\end{equation}

In this method, we fit the sky fiber CCF with a Gaussian to measure both the continuum level and RV position of the scattered sunlight relative to the target star. Since the sky spectrum is approximately at the barycentric velocity, the CCF center can be fixed or heavily constrained for the peak fitting process. The estimated sky brightness is used to scale a noiseless model sky spectrum at the correct flux level. In this idealized scenario, we use the same synthetic spectrum that was used to generate the injected sky spectrum. In reality, this model should be drawn from a library of high SNR observations of twilight sky at that particular site, preferably observed through the science fiber itself to facilitate better LSF matching.

The model sky spectrum is subtracted from the science fiber spectrum, and a final CCF of the corrected target spectrum is generated. The results of this correction are shown in Figure \ref{fig:corrected} (right). This technique also reduces the solar contamination effect considerably, bringing the worst RV errors down from $\sim$100~\ms~to $\sim$1~\ms. It does not, however, work as well as simple subtraction of the sky fiber CCF, since the extra steps of measurement, translation, and interpolation of spectra can introduce small errors.

This technique could be particularly useful for configurations where the sky and science fibers have different geometries or spectral resolutions, where direct CCF subtraction is not appropriate. Using the sky spectrum CCF to scale the template solar spectrum, rather than the recorded sky spectrum itself, may be advantageous for low-sky background levels where the continuum is difficult to identify. However, there is an additional burden in ensuring that the model spectrum matches the true sky spectrum based on site, airmass, moon phase, and other observing variables.

\subsection{Limitations of the Sky Fiber}

Both of the solar contamination correction techniques described above rely on the use of a simultaneous sky fiber. The CCF subtraction method further relies on the similarity of the science and sky fibers (or the ability to transform the LSF and chromatic variability between the two). However, there may be times when a sky fiber is not available (in certain instruments one has to choose between calibration light or sky illumination in the simultaneous fiber), or has contamination issues itself.

There are two main locations that introduce contamination into the sky fiber. The first occurs at the port adapter fiber head (Figure \ref{fig:fiber_head}), and is caused by the wings of the stellar PSF extending into the area covered by the sky fiber (22'' separation). For a $V_{\rm mag}=12$ star and a sky brightness of 17~mag/arcsec$^2$, the sky fiber receives 0.001\% extra contaminating flux; as sky dims to 21.45~mag/arcsec$^2$ (nominal for WIYN dark time at zenith), this increases to  0.08\%. Irrespective of instrument, some contamination at the port fiber head is unavoidable given the desire to sample the same (or adjacent) patch of sky. This effect is naturally mitigated in better seeing, but the fact that it varies with observing conditions adds further complexity to modeling efforts.

The second, and more serious, source of concern is cross-contamination between fiber traces on the detector, caused by low-level wings of the cross-dispersion profile. Leaking of starlight into the sky fiber of the same echelle order (intra-order contamination), expected to be between $10^{-3}-10^{-5}$ levels, is not hugely problematic in the CCF correction method, since it essentially results in a slightly depleted primary CCF peak. However, attempts to gauge sky brightness from the sky fiber are strongly affected, as a function of stellar brightness. On the red end of the NEID bandpass, inter-order contamination becomes much more significant, as the  spatial separation between the sky fiber and the neighboring calibration order diminishes to $\sim$7 pixels (compared to $\sim$80 pixels at the blue end). Attempted measurements of the sky fiber spectrum get quickly overwhelmed by light from the calibration source (laser frequency comb, etalon, or emission lamp). To fully utilize the sky fiber for NEID, calibration light will need to be filtered in the redder orders, although we will also test modeling and removal of cross-contamination on real data.

\begin{figure*}[t]
    \begin{center}
        \includegraphics[width=0.9\textwidth]{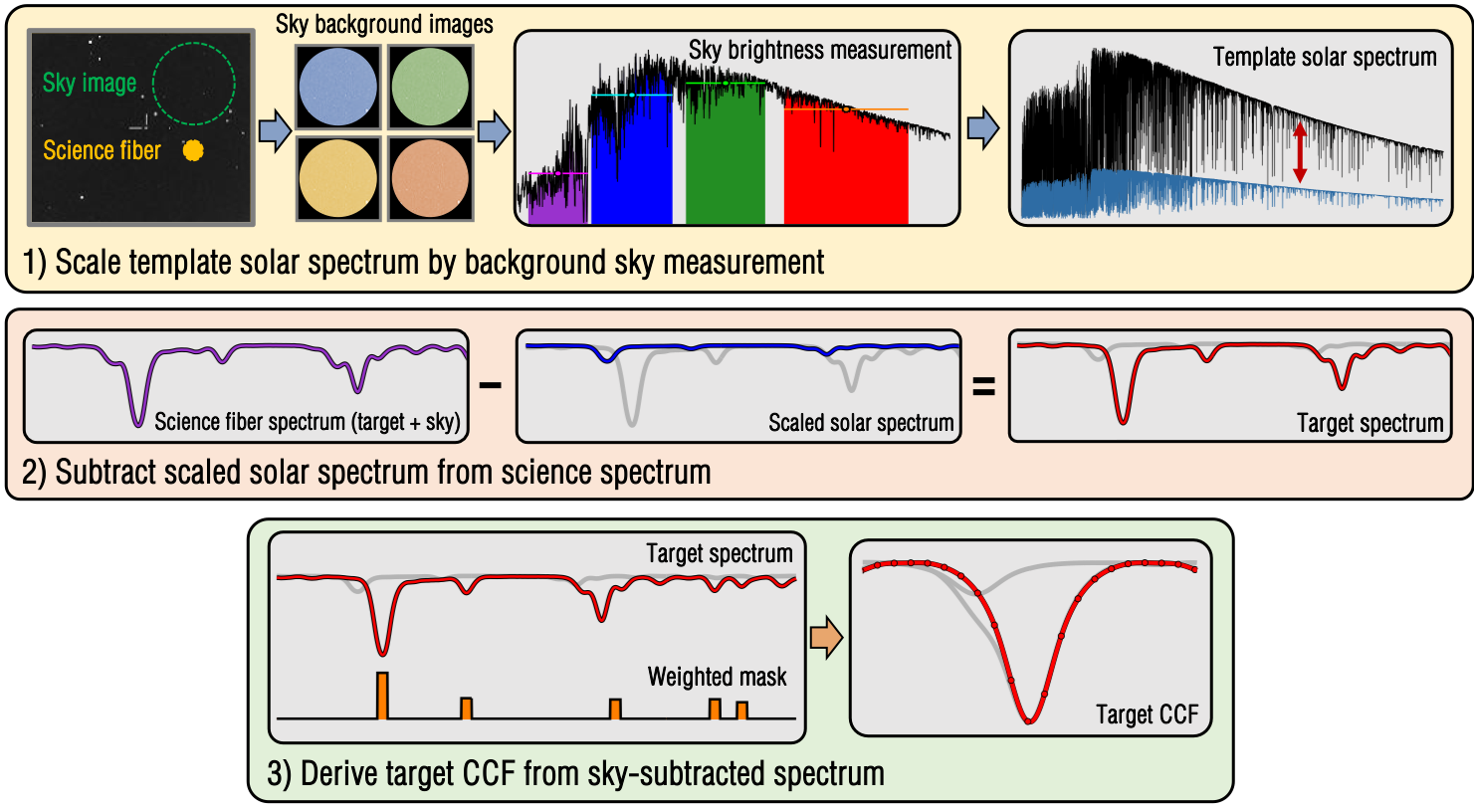}
        \caption{Overview of the simultaneous broadband correction technique, as described in \S\ref{sec:broadband}. An image of the sky in one or more photometric bands is used to scale a solar template spectrum as a proxy for the background contamination. The scaled solar spectrum is then subtracted from the science spectrum, leaving only the target spectrum (middle). The final target radial velocity is then computed from the target CCF (bottom). In the case of NEID, a coherent fiber imaging bundle is used to sample the background sky in the vicinity of the science target. Note that the spectrum and CCF continuum levels are matched for illustrative purposes only -- in reality, the sky fiber has a much lower flux level.}
        \label{fig:ccf_cfb}
    \end{center}
\end{figure*}

\begin{figure*}[ht]
    \begin{center}
        \includegraphics[width=1\textwidth]{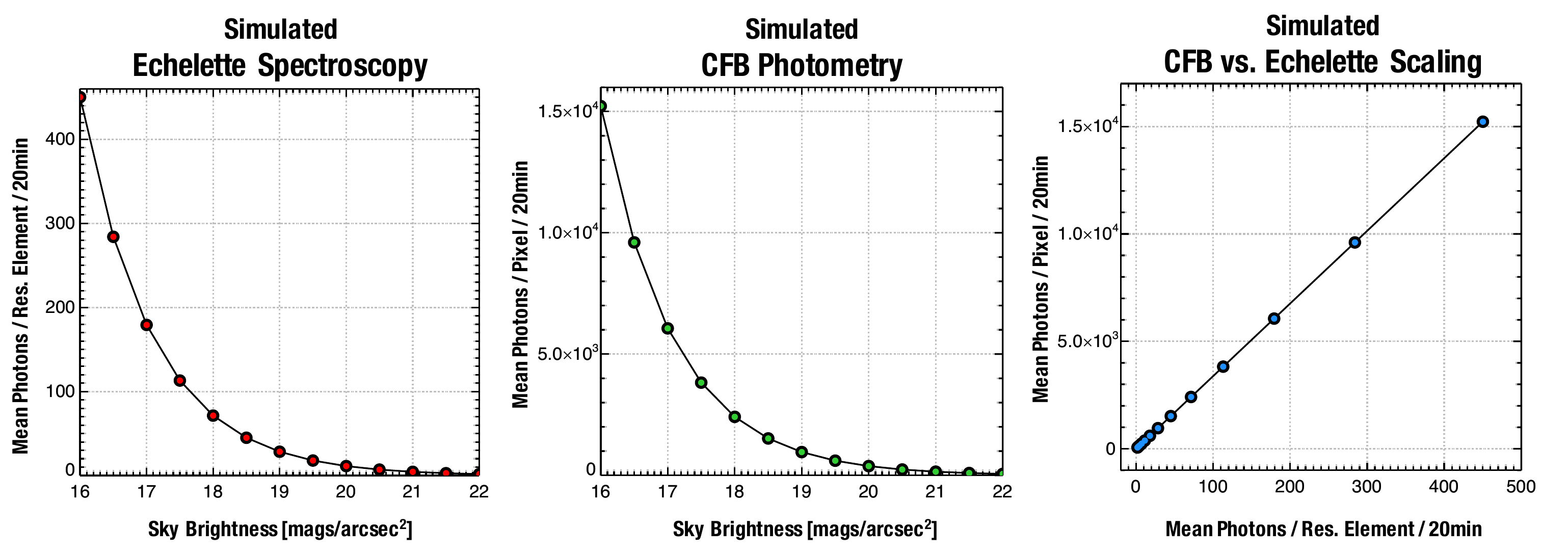}
        \caption{Predicting the relationship between the sky fiber spectrum and broad-band images from the CFB. \textit{Left:} Mean photons per resolution element in the NEID sky fiber spectrum, for a 20 minute exposure, as a function of sky brightness. \textit{Middle:} Number of photons per pixel in a CFB V-band image of the sky for the same exposure time. For the sake of direct comparison we are not applying the saturation limit of the detector. \textit{Right:} Relation between the counts in the broadband CFB image and the flux in the spectrum, showing a clean linear scaling that can be easily extended to fainter skies.}
        \label{fig:sky_prediction}
    \end{center}
\end{figure*}

\begin{figure*}[ht]
    \begin{center}
        \includegraphics[width=1\textwidth]{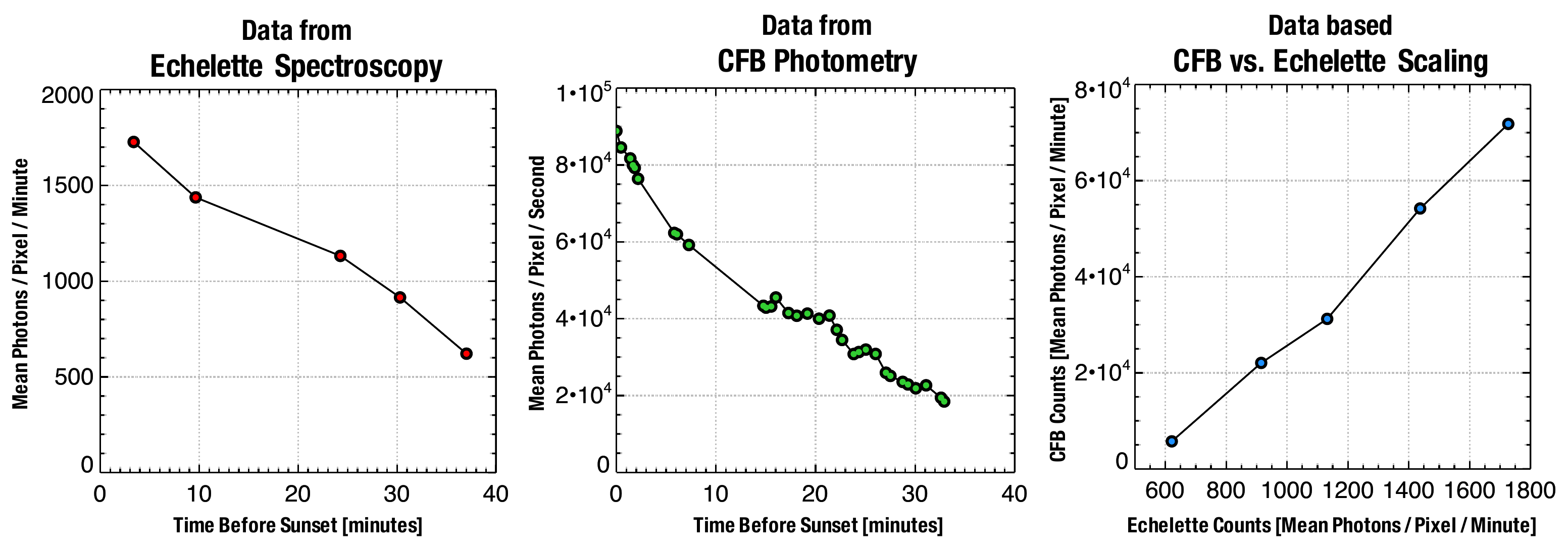}
        \caption{On-sky proof of concept that a relationship can be established between the sky fiber spectrum and broadband images from the CFB, as predicted in Figure \ref{fig:sky_prediction}. Data was taken as the sun set, to collect exposures at different sky brightnesses. \textit{Left:} Mean photons per extracted pixel in the echelette sky spectrum, all exposures have been scaled to 1 minute. \textit{Middle:} Number of photons per pixel in a CFB V-band image of the sky, all exposures scaled to 1 second. \textit{Right:} Relation between the flux in the spectrum and the broadband CFB image, proving a clean linear scaling that can be extended to fainter skies. Note that the CFB counts are binned to 1 minute for direct comparison.}
        \label{fig:sky_data}
    \end{center}
\end{figure*}

\begin{figure*}[ht]
    \begin{center}
        \includegraphics[width=0.7\textwidth]{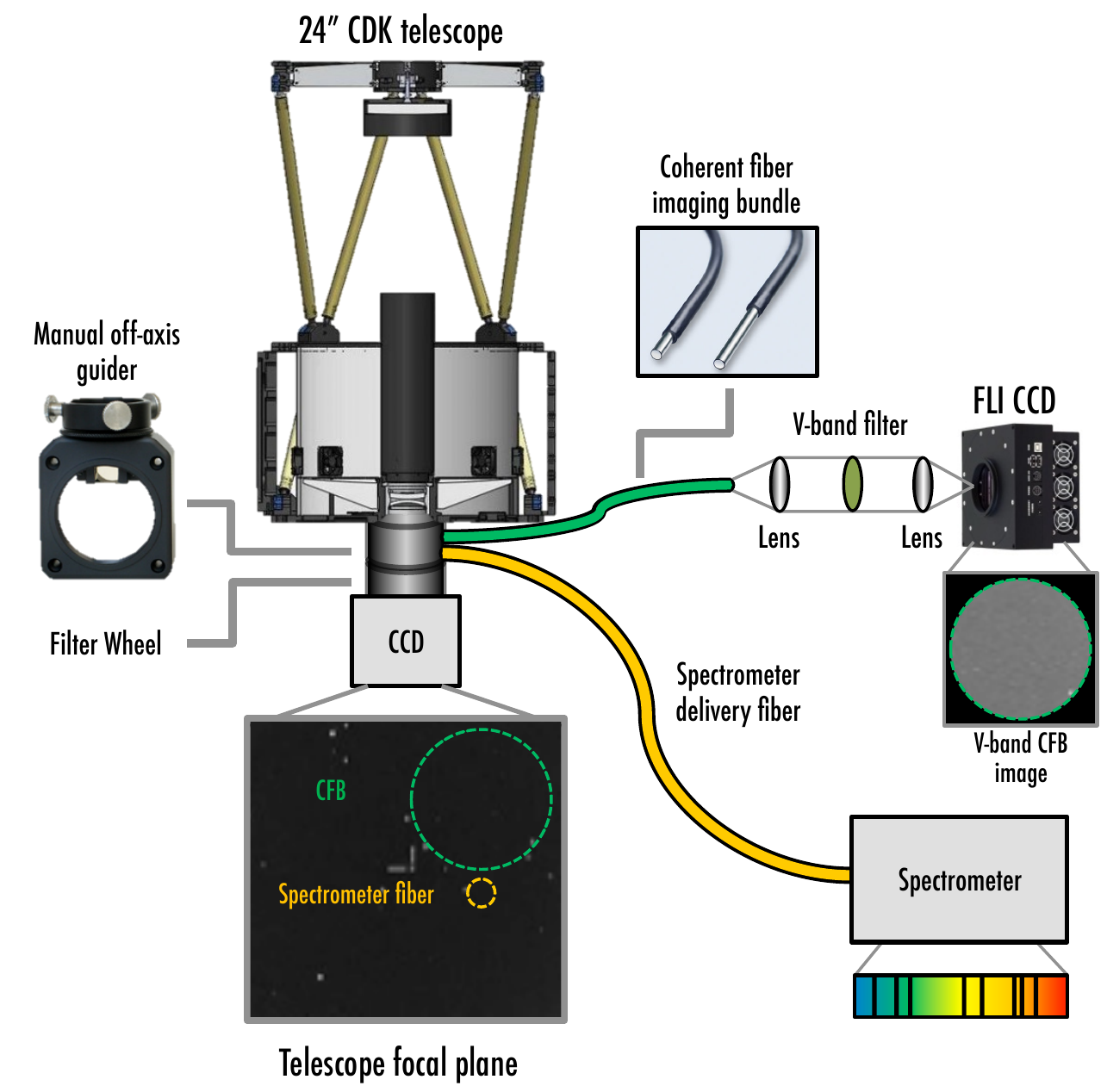}
        \caption{On-sky instrumental setup for scattered sunlight measurement. A small fraction of light collected by the 24-inch PlaneWave is diverted to a secondary image plane formed at the focus of the manual off-axis guider assembly (MOAG). A coherent fiber bundle (CFB) and 300 $\mu$m fiber are placed at the focus of the MOAG. The CFB output is imaged onto a FLI CCD, while the 300 $\mu$m fiber is inserted into a compact high resolution spectrometer.}
        \label{fig:on_sky_setup}
    \end{center}
\end{figure*}

\begin{figure*}[ht]
    \begin{center}
        \includegraphics[width=0.8\textwidth]{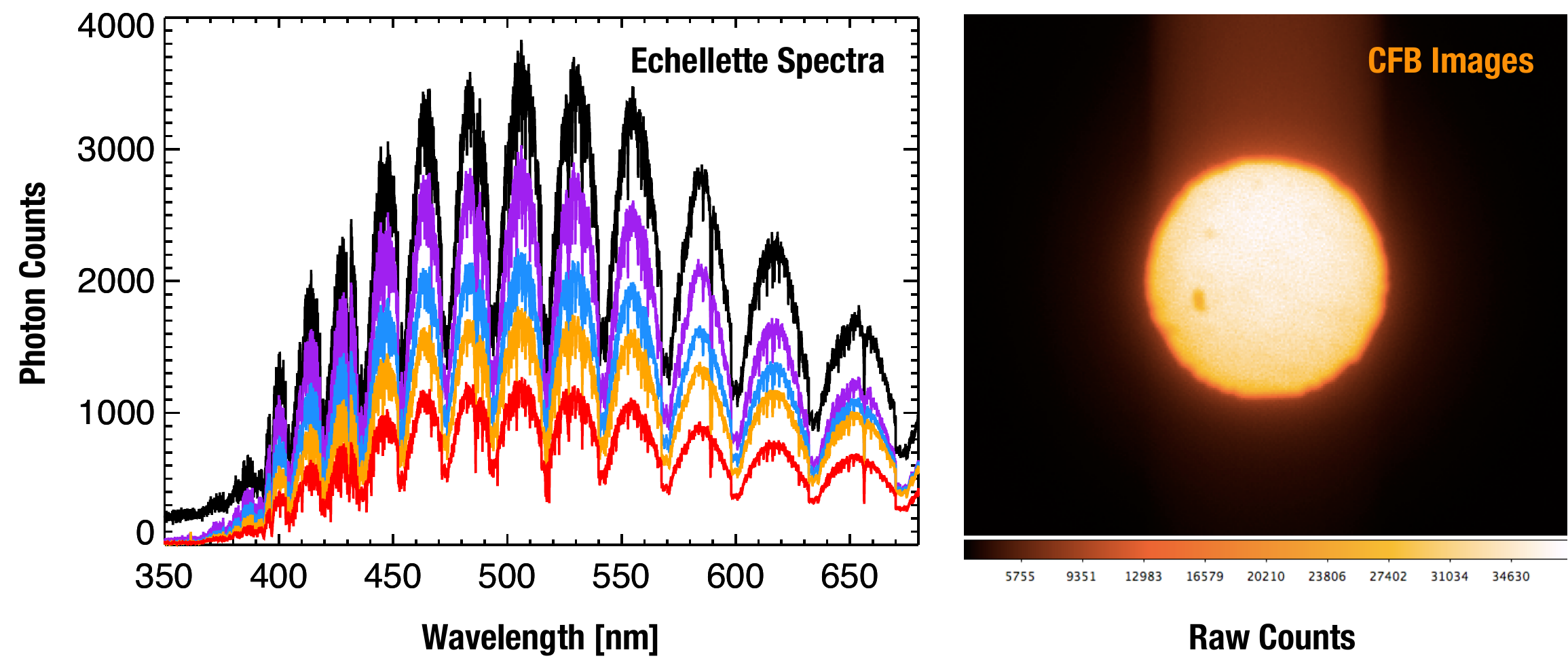}
        \caption{Simultaneous spectra and images taken of the sky as the sun set. \textit{Left:} Moderate resolution (R$\sim$10,000) optical spectrum of the sky taken with a small commercial echelette spectrometer calibrated with an Hg lamp. \textit{Right:} A coherent fiber bundle (CFB) imaged on an FLI CCD.}
        \label{fig:cfb_data}
    \end{center}
\end{figure*}

\subsection{Simultaneous Broadband Correction}
\label{sec:broadband}
Given the low flux levels and concerns of contamination in the sky fiber, simultaneous broadband measurements can provide important redundancy and might become the only viable option for the darkest skies. In this method, the level of scattered solar irradiance is estimated by sampling the background sky via broandband imaging, preferably in reasonably close proximity to the target object. This yields an independent estimate of the solar contamination, and does not rely on the potentially noisy high resolution sky spectrum. An overview of the technique is shown in Figure~\ref{fig:ccf_cfb}.

Coherent fiber bundles (CFBs), with several thousand fiber cores, are flexible image guides that enable large fields of view relative to typical spectrograph fibers, and are often used in astronomical observatories for target acquisition, guiding, and maintaining focus \citep{Ramsey:2003, Newman:2004,Smee:2013, Yan:2016}. For NEID, three CFBs will surround the science fiber in the port adapter fiber head. This design is itself adapted from the Habitable-Zone Planet Finder \citep[HPF,][]{Metcalf:2019} fiber head design \citep{Kanodia:2018}, which in turn is modeled on a design proposed by Phillip McQueen (private communication) for the Hobby-Eberly telescope's upgraded high-resolution spectrometer (HRS).  

Figure \ref{fig:fiber_head} shows the relative placement of the CFBs with respect to the science and sky fibers in NEID, as well as a magnified image of a single CFB. The CFBs (6.3'' on sky) will be used to regularly calibrate the port tip/tilt module and triangulate the precise center of the science fiber in the focal plane. This triangulation system provides a robust measurement of the location of target star relative to the science fiber. All CFBs will be imaged onto a single commercial FLI MicroLine CCD, and continue to collect data throughout the night. The CFBs will thus be gathering large amounts of sky light simultaneously with science exposures. The presence of three bundles creates additional redundancy in case of bright stars drifting into their fields of view. While contamination from stellar light at the port adapter fiber head can also affect the CFBs (15'' separation from science fiber), they are immune to concerns of cross-contamination on the detector, making them an important complement to the sky fiber.

The CFBs provide a new and potent avenue for the correction of solar contamination using simultaneous broadband sky imaging. By analyzing CFB images in tandem with the science fiber observing relatively bright sky, a scaling relation can be set up between the CCF continuum and the absolute sky brightness in the V-band. Figure \ref{fig:sky_prediction} shows the simulated relative efficiency in sky flux collection between the NEID fiber and each CFB, and the predicted linear scaling relation between the two. This scaling can be extended to darker skies, allowing the option to circumvent the sky fiber when it becomes less reliable. The sky brightness level measured from the CFB can consequently be used to scale a noiseless model sky spectrum and subtract it from the science spectrum as described above. 

One caveat to note here is that the CFB image is not dispersed and hence cannot communicate chromatic variations of the sky SED as weather changes during the observation. While the sky fiber technically contains this information, in both previous correction methods we ultimately measure the CCF, which collapses wavelength in favor of SNR, thereby forsaking this knowledge as well. Chromatic yet efficient sky measurements would require, for example, different filters on the three CFBs, which are not feasible with the current NEID port design. The CFBs also make it difficult to distinguish between fluctuations in scattered sunlight versus airglow or city lights; for NEID we try to limit these issues with custom filters (\S\ref{sec:neidcfb}).

\subsubsection{On-sky Demonstration}
As proof of concept, we conducted an on-sky experiment at Penn State to show that the broadband image from the CFB can, in fact, be used as a reliable proxy for the sky fiber. A commercial 24 inch PlaneWave CDK was used to feed a dedicated optical fiber that directly measured the scattered night sky spectrum. We recorded spectra from evening to after dusk, mimicking the change in sky brightness for a variety of observing conditions and lunar separations. Figure~\ref{fig:on_sky_setup} shows our on-sky measurement scheme using the 24 inch telescope. A fraction of light from the telescope pupil is picked off using an off-axis mirror within the manual off-axis guider (MOAG) assembly. The MOAG produces an image of the night sky that simultaneously illuminates a 300 $\mu$m optical fiber and a 1.4 mm coherent fiber imaging bundle (CFB) cable. The 300 $\mu$m optical fiber is fed into a small commercial echelette spectrometer, which produces a cross-dispersed, moderate resolution (R$\sim$10,000) optical spectrum (400 -- 600 nm) of the night sky. The output of the CFB is imaged onto a 1024 $\times$ 1024 FLI CCD using a simple lens relay system that includes a Johnson V-band filter. Figure~\ref{fig:cfb_data} shows the extracted spectra and a sample image of the CFB. The FLI CCD was experiencing a shutter lag during these exposures due to its orientation with respect to the telescope (a known issue with these cameras), leading to streaking in the images. This was corrected out using darks and an interpolation algorithm to estimate the underlying flux from the shutter bleed.

The goal of this experiment was to explore the feasibility of using integrated V-band flux as an independent method for estimating the solar contamination level. Ideally, the broad-band measurement from the CFB can be used to reliably scale a template solar spectrum, which would then yield an estimate of the absolute solar contamination level. To test the validity of this technique we compare the measured spectrum from the optical fiber, under a variety of different pointings and conditions, to the scaled solar spectrum from the broadband CFB image (Figure \ref{fig:sky_data}). This follows the behavior predicted in Figure \ref{fig:sky_prediction}, with time as a surrogate for sky brightness. Flux scaling of the CFB image thus has the advantage of providing an independent verification of contamination levels, as well as potentially yielding a significantly higher signal-to-noise measurement of the solar contamination than the recorded high resolution spectrum from the single optical fiber.

\section{Building Correction Methods into NEID Design}
\label{sec:neidcfb}

\begin{figure*}[ht]
    \begin{center}
        \includegraphics[width=1\textwidth]{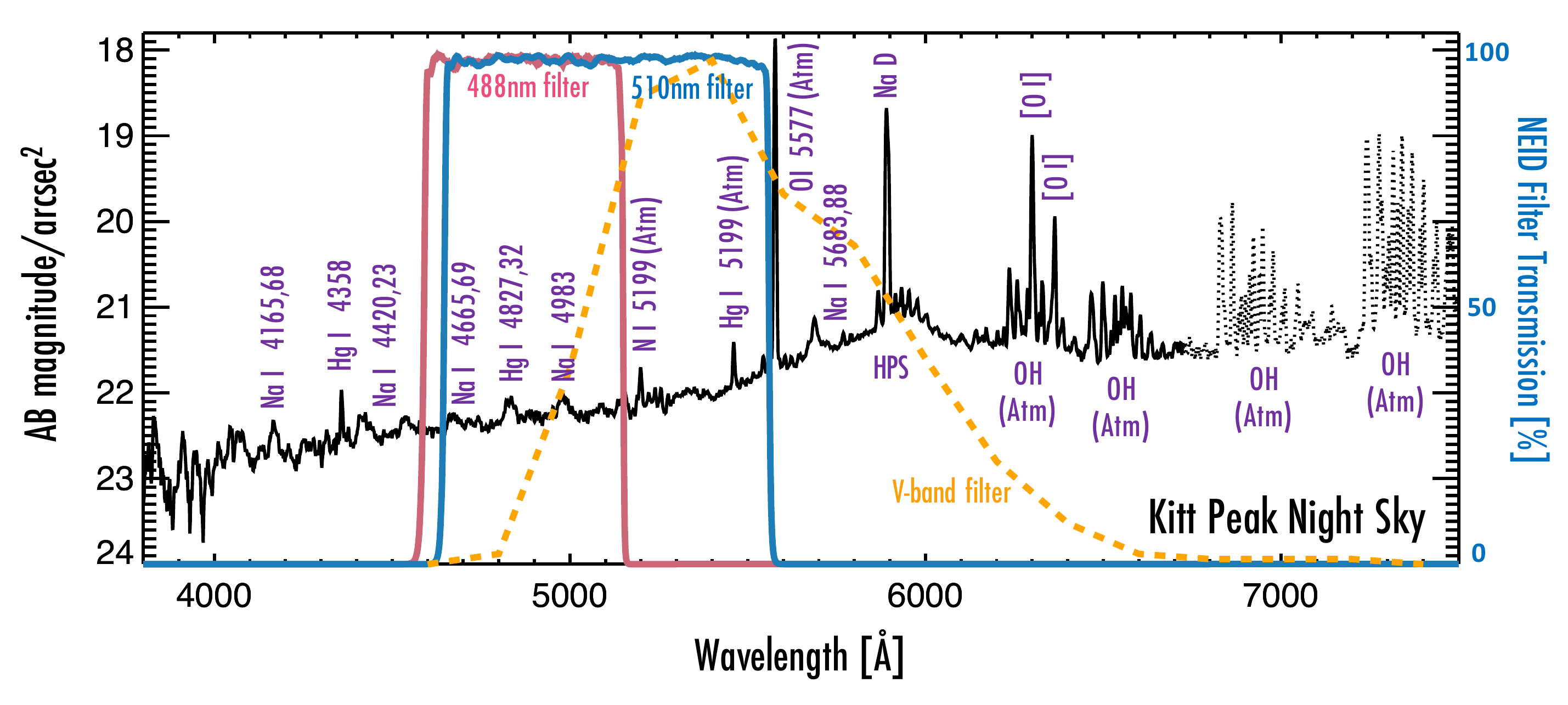}
        \caption{Spectrum of the night sky at zenith above Kitt Peak from \cite{Massey:2000}, overlaid with the NEID CFB filter curves and a typical V-band filter. Emission features from both natural and human-made sources are highlighted. High-pressure sodium (HPS) lamps cause the broad feature extending from $\sim 5500 - 6500$~\AA. The two filters from Semrock, centered on 488nm and 510nm, were carefully selected to avoid bright (and variable) emission lines that could confound measurement of overall sky brightness, as well as good out of band suppression of the light. Filter transmission begins after the HgI 4358 feature, and cuts off before the strong OI feature at 5577~\AA, and the Na D lines.}
        \label{fig:kpnosky}
    \end{center}
\end{figure*}

NEID includes a dedicated CFB imaging camera, primarily used to aid in optical registration of the telescope focal plane with the NEID science and sky fiber \citep{Logsdon:2018, Schwab:2018}. The CFBs are mechanically mounted onto the same fiber connector as the NEID science and sky fibers and, when not in use as target registration fiducials, will provide independent samples of the background sky. 

The outputs of the CFBs are imaged through a filter wheel onto a commercial FLI CCD package using a pair of telecentric imaging lenses. In addition to some standard neutral density filters for the target registration function of the CFBs, we include high performance off the shelf (Semrock) spectral filters for monitoring sky brightness. Instead of limiting the sky measurement to a standard V-band filter (which is included for target centering via CFB triangulation), we have two tailored filters centered at 488nm and 510nm to ensure bright airglow lines are avoided in the final recorded images (Figure \ref{fig:kpnosky}). We are particularly careful to avoid the strong OI feature at 5577 \AA, and the Na I D lines, and ensure good suppression outside the desired filter window. Avoiding highly variable, bright airglow lines is essential for ensuring the CFBs are tracing scattered solar contamination, and not variability due to other atmospheric effects. 

In addition to the CFBs, NEID also includes a chromatically dispersed exposure meter to aid in performing accurate barycentric corrections. While the exposure meter does simultaneously record low resolution spectra from all three NEID fibers, it only contains a very small fraction ($<1$\%) of the light in the main spectrograph, leading to extremely small levels of flux in the sky fiber trace. In combination with the fact that it is read out very fast to carefully track the flux accumulation during an exposure, the exposure meter is not a viable source of additional sky brightness monitoring for NEID. 

\section{Discussion}
\label{sec:discussion}
While our calculations show that RV errors due to scattered solar contamination may be mitigated using several different techniques, there are a certain site-based and instrument-specific effects that we exclude for the sake of generalization. As listed below, these must all be considered when applying solar contamination correction to real observations, and can add further complexity to the process.

\begin{enumerate}

\item Any sky fiber will be spatially separated from the science fiber in the spectrometer focal plane, leading to different aberrations and a fundamentally different PSF compared to the science fiber. The degree to which this will affect different correction techniques will vary based on the spectrometer design, though we aim to probe this with NEID in the near future.

\item We have assumed the science and sky fibers have identical (or perfectly characterized) throughputs including focal ratio degradation. This will also be system-specific, since fibers naturally have different transmission properties depending on manufacturing.

\item The sky is assumed to be uniform, both spatially and spectrally, between the science and sky fibers, which may not be the case depending on angular separation on-sky.

\item The SED of the sky contamination is assumed to be identical to the Sun, though in reality atmospheric conditions will modulate the solar spectrum chromatically to some degree \cite[e.g.,][]{Brine_1983}.

\item We assume that the sky fiber does not suffer significant contamination from the science or calibration fiber. While this is instrument dependent, it is unlikely to be strictly true for any instrument. The wings of stellar PSF overlapping with the sky fiber at the telescope fiber injection unit, scattering from the cross-disperser, scattering from optics like the primary echelle grating (e.g., the Rowland Ghost), and intra- and inter-order cross-contamination between fibers on the detector will all contribute to stellar or calibration light contamination in the sky fiber.

\end{enumerate}

\section{Conclusion}
\label{sec:conclusion}

We present an analytical study estimating the Doppler RV error due to scattered sunlight contamination in RV observations using next generation spectrometers. This subtle effect, while low in absolute amplitude, can begin to inhibit high precision Doppler measurements at the sub-{\ms} level if not explicitly corrected. Brighter stars are naturally better inured against typical levels of scattered solar contamination than fainter stars, which show orders of magnitude larger RV errors. 

To begin with, we show that this effect \textit{must} be acknowledged when scheduling observations, and that one cannot observe in all conditions of sky brightness, moon separation, and moon phase, without considering the solar contamination error. If overlooked, this error source can easily grow to several \ms~~in magnitude and dominate the meager error budgets of next-generation instruments. We also demonstrate that the presence of a sky fiber is ideal, particularly if it has identical spectral properties (resolution, PSF, throughput) to the science fiber. If no sky fiber is available, or the sky is too faint to reliably measure from the sky fiber, simultaneous broadband sampling of the sky shows great promise. Lastly, we emphasize that some level of sky correction is going to be an absolute necessity in the limit of faint targets (e.g., for follow-up of TESS targets with $V_{\rm mag} > 12$) or highest desired precision (potential small rocky planets around bright stars). 

We present three prospective correction methods: (1) subtracting the CCF of a dedicated sky fiber with the same spectral properties (resolution, throughput) as the target fiber, (2) using the sky fiber CCF to subtract a scaled model solar spectra and (3) using simultaneous broadband measurements of the background sky to scale a template solar spectrum, using CFBs in the case of NEID. While simple CCF subtraction works best in the case of the sky fiber having the same spectral properties as the science fiber, subtracting model spectra presents a comparable alternative. The latter technique has the added benefit of not requiring the same PSF properties for both the science and sky fibers.

However, our presentation of the sky fiber usage here is in many ways idealized, since we do not account for complex effects like cross-contamination between fibers, and other sources of temporal variability between the sky and science fiber. As an alternative to using the sky fiber, we present a third option of using a broadband source like the NEID coherent fiber bundles, with superior collecting power, to gain a coarse sky brightness measurement instead. The CFBs in NEID are installed primarily for target acquisition, but naturally observe sky during all science observations. Recognizing the potential for a high-SNR proxy for the sky fiber early on, we have built towards this new functionality, and anticipate demonstrating the required suppression of solar contamination with NEID.

Minimizing the impact of solar contamination to very low levels requires a combination of strategies. Careful planning of observations taking into account moon phase, radial velocity and barycentric correction of the star, hardware solutions like sky fibers and CFBs to accurately measure the level of sky contamination, and algorithms to perform accurate corrections. Our intent in this paper is to present and model the deleterious impact of solar contamination, and show that it can, at least in principle, be corrected to well below 10 cm/s using a combination of strategies. Future experiments with the publicly available NEID radial velocity platform on the 3.5m WIYN telescope will let us truly probe the effectiveness of these strategies in the face of real on-sky and instrument systematics, as well as explore the path to mitigating these issues further as we gather longer baselines of data at extreme precision.

\facilities{DAVEY:0.6m}

\acknowledgments
On-sky tests were conducted with the Planewave CDK 24'' Telescope operated by the Penn State Department of Astronomy \& Astrophysics at Davey Lab Observatory. NEID is funded by NASA through JPL by contract 1547612. This work was partially supported by funding from the Center for Exoplanets and Habitable Worlds. The Center for Exoplanets and Habitable Worlds is supported by the Pennsylvania State University, the Eberly College of Science, and the Pennsylvania Space Grant Consortium. This work was performed by SPH  [in part] under contract with the Jet Propulsion Laboratory (JPL) funded by NASA through the Sagan Fellowship Program executed by the NASA Exoplanet Science Institute. This work was supported by NASA Headquarters under the NASA Earth and Space Science Fellowship Program through grant NNX16AO28H. This research has made use of NASA's Astrophysics Data System Bibliographic Services.

\bibliography{ms}{}
\bibliographystyle{aasjournal}

\end{document}